\documentclass[journal=jpcbfk,manuscript=article]{achemso}

\usepackage[version=3]{mhchem} 
\usepackage{comment}

\author{Jiyong Cheon}
\affiliation{Department of Physics, Ulsan National Institute of Science and Technology, Ulsan, Republic of Korea}

\author{Joonwoo Jeong}
\affiliation{Department of Physics, Ulsan National Institute of Science and Technology, Ulsan, Republic of Korea}
\email{jjeong@unist.ac.kr}
\phone{+82-52-217-2155}

\title[An \textsf{achemso} demo]{Solvent isotopic effect on the phase transition of lyotropic chromonic liquid crystals: Heavy water makes mesogens less charged}

\abbreviations{IR,NMR,UV}
\keywords{isotopic effect, chromonic liquid crystal, phase transition, deuterium}

\begin{document}

\begin{abstract}
The interplay among solute and solvent molecules in lyotropic mesophases governs their physicochemical properties, such as phase behaviors and viscoelasticity. 
In our model system, a lyotropic chromonic liquid crystal (LCLC) made by disodium cromoglycate (DSCG), charged plank-like molecules self-assemble to form elongated aggregates via non-covalent attractions in water (H$_{2}$O). 
The aggregates align to exhibit liquid crystalline phases: nematic and columnar phases. 
Here, we report the isotopic effect on the phase behavior of the LCLC when D$_2$O is substituted for H$_2$O. 
D$_2$O-DSCG exhibits higher nematic-to-isotropic phase transition temperatures than H$_2$O-DSCG. 
X-ray scattering reveals considerably longer inter-aggregate correlation lengths in D$_2$O-LCLCs. In contrast, the other microstructural properties, such as inter-aggregate distances and intra-aggregate correlation lengths, remain almost the same. 
Our $^{23}$Na FT-NMR measurement reveals that D$_2$O-DSCG aggregates are less charged with more counter-ions, Na$^+$, bound to them than H$_2$O-DSCG aggregates. 
Weaker electrostatic repulsion between aggregates may stabilize the nematic phase, and this solvent isotopic effect may generally apply to diverse aqueous lyotropic mesophases with electrostatic interactions.
\end{abstract}

\section{Introduction}
Lyotropic liquid crystalline mesophases form when solute molecules or colloids interact to make partially ordered structures in a solvent, depending on concentration and temperature~\cite{neto2005}. 
The solvent-mediated inter-molecular or inter-colloidal interactions are responsible for self-assembled structures, which can be hierarchical, ranging from micelles to nematic alignment, lamellae, and gyroid~\cite{Vroege_1992, dierking2017, Marco2017, Davidson2018}. 
These hierarchical structures and underlying interactions govern their phase behaviors and macroscopic properties, such as viscoelasticity and birefringence~\cite{Hongladarom1996, Zhou2014, zhang2018}, which are of keen interest in applications. 

Various techniques spanning optical microscopy, X-ray/neutron scattering, and NMR spectroscopy have been employed for phase and structure characterization of mesophases, and the isotopic substitution has played a pivotal role.
Notably, deuterium(D) substitution for hydrogen(H) has been widely applied because of their physicochemical similarity but significant differences in the magnetic moment and neutron scattering cross-section~\cite{jeffries2021}.
For example, 2D-NMR and mass spectrometry investigated deuterated proteins' folding dynamics and structure~\cite{Englander1992, Konermann2011} or a deuterated detergent examined the membrane protein structure by small angle neutron scattering~\cite{conn2021, Golub2022}. 
Substituting aqueous solvent, \textit{i.e.},  D$_2$O for H$_2$O, also revealed the structural information of biological systems, such as proteins, ribosomes, and viruses~\cite{BJacrot1976}.
Although it is often assumed these isotopic substitutions minimally affect systems, such as polymers, proteins, and liquid crystals, there do exist isotopic effects altering their phase behaviors and structures~\cite{Barron2011, makhatadze1995, Barbara1982, Middleton2007, kostko2005, Shvartzman2009}.
Scrutinizing the isotopic effect not only matters in the phase and structure characterization but also deepens our understanding of the lyotropic systems, \textit{e.g.}, disclosing the role of H-bonds in the formation of worm-like micelles~\cite{Barron2011} or micellization of block copolymers~\cite{Shvartzman2009} thanks to stronger D-bonds.

This work investigates the solvent isotopic effects in a lyotropic chromonic liquid crystal (LCLC) called disodium cromoglycate (DSCG).
The plank-like molecules dissolved in water form rod-like aggregates via non-covalent face-to-face stacking, and the charged aggregates align to form nematic or columnar phases according to temperature and concentration~\cite{Nastishin2004, Lydon2011}.
LCLCs' phase behaviors have been studied thoroughly, according to the concentration, temperature, pH~\cite{Zhang2016}, counterions~\cite{Suk2008}, external fields~\cite{Nastishin2018, LI2022}, impurities, and additives~\cite{Yu1982, kostko2005, Park2008, Park2011, Yamaguchi2016, Hyesong2019, Eun2020}. We focus on rationalizing the solvent isotopic effect reported in Kostkov \textit{et al.}~\cite{kostko2005}, where D$_2$O substitution for H$_2$O increases the phase-transition temperature, and reveal that D$_2$O makes the aggregates less charged, stabilizing the nematic phase.

\section{Method}

\textbf{Sample Preparation and Measurement of Phase Transition Temperatures} We used cromolyn sodium salt (DSCG, Sigma-Aldrich) with 95\% purity to make the liquid crystal samples in the experiments. We dissolved the DSCG powder into deionized H$_2$O and deuterium oxide (99.9\% D purity, Sigma-Aldrich), respectively, varying the molar fractions exhibiting a fully nematic phase at room temperature.

We employed polarized optical microscopy (POM) to observe the phase behavior of the specimens controlling their temperatures. 
A 100~$\mathrm{\mu m}$-thick sandwich cell filled with DSCG solution was sealed by epoxy glue to minimize the evaporation of the solution. 
Then, we observed the cell under a polarized optical microscope (BX53-P; Olympus) equipped with a temperature-controlled stage (T95-PE120; Linkam Scientific Instruments), a color CCD camera (INFINITY3-6URC; Lumenera), a quasimonochromatic illumination (wavelength = 660 nm, FWHM = 25 nm) derived from an LED lamp (LED4D067; Thorlabs), and a 10x dry objective lens for a large field of view. 
Then, increasing temperature from the room temperature at the rate of 0.5$^\circ$C/min, we recorded the temperature where isotropic domains started to appear within the fully nematic phase.

\textbf{X-ray scattering measurement and data analysis.}
We conducted wide-angle X-ray diffraction (WAXD) and small-angle X-ray scattering (SAXS) experiments at the PLS\MakeUppercase{\romannumeral 2} 6D UNIST-PAL beamline of the Pohang Accelerator Laboratory (PAL), Republic of Korea. 
X-ray energy was 11.564 keV, and the WAXD and SAXS covered the $q$-range from 0.11787 to 3.36527~\AA$^{-1}$ and 0.05 to 0.2~\AA$^{-1}$, respectively.
All measurement data were acquired from the DSCG solutions in borosilicate rectangular capillaries (Vitrocom) having a 100 $\mathrm{\mu m}$ path length at 23.0 $\pm$ 1$^\circ$C.
We controlled the X-ray exposure time, \textit{i.e.}, 10 s, to minimize the X-ray damage to the specimen; See Fig. S4 for the effect of X-ray exposure to the DSCG specimen.
Following Eun \textit{et al.}\cite{Eun2020}, we fitted angle-averaged intensity profiles using the Lorentzian function with a linear background after subtracting the background signal measured from the empty capillary.
Then, we retrieve the average inter-aggregate distance $D$ and correlation length $\xi_D$, and the average inter-molecular distance $d$ and correlation length $\xi_d$. Specifically, the position and full width at half-maximum of the peak in the small-angle region around 0.12 \AA$^{-1}$ give D and $\xi_D$, respectively. The ones of the wide-angle peak give d and $\xi_L$. See the Supporting Information for the X-ray scattering data.

\textbf{$^{23}$Na FT-NMR measurement and quadrupole splitting analysis} The $^{23}$Na FT-NMR experiment was conducted using a high-resolution 400 MHz FT-NMR spectrometer (MODEL; Bruker). The DSCG specimens were prepared in a 5 mm NMR tube (541-PP-7; Wilmad), while a separate 5 mm NMR tube (WGS-5BL; Wilmad) containing a deuterated solvent, dimethyl sulfoxide-d$_6$ (Sigma-Aldrich), was prepared as a chemical shift reference. The NMR tube with the chemical shift reference was coaxially inserted into the liquid crystal sample tube. We used MNOVA (Mestrelab Research) to process the data acquired by 4096 scans, including phase correction and baseline subtraction. See Fig. S3 for the details of FT-NMR data and how we applied three-Lorenztian peak fittings to estimate the quadrupole splitting. 

\section{Results and discussion}

\begin{figure}[htp!]
\centering 
\includegraphics{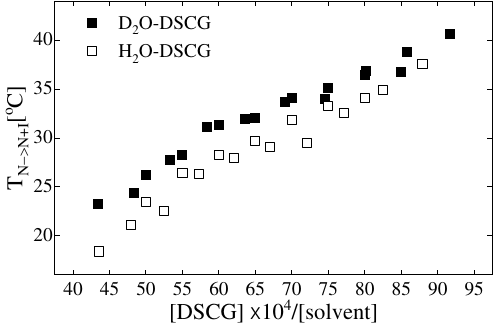}
    \caption{The phase transition temperature according to the concentrations of DSCG in D$_2$O and H$_2$O. The filled and empty squares correspond to D$_2$O-DSCG and H$_2$O-DSCG, respectively. The horizontal axis is $10^4$ times shown as the molar ratio of DSCG molecules to solvent molecules.}
    \label{fig:phase diagarm}
\end{figure}

As shown in Fig.~\ref{fig:phase diagarm}, the nematic-to-coexistence transition temperature of DSCG dissolved in D$_2$O (D$_2$O-DSCG) is approximately 5$^\circ$C higher than the transition temperature of H$_2$O-DSCG. 
See Methods for how to measure the transition temperature.
Namely, D$_2$O-DSCG forms a more stable nematic phase at the same concentration, which is represented as the number of DSCG molecules per the number of solvent molecules (H$_2$O or D$_2$O) in Fig.~\ref{fig:phase diagarm}. 
For instance, 60 DSCG molecules in $10^4$ solvent molecules correspond to 14.57 wt/wt\% for H$_2$O-DSCG and 13.31 wt/wt\% for D$_2$O-DSCG.

\begin{figure}[htp!]
\centering 
\includegraphics{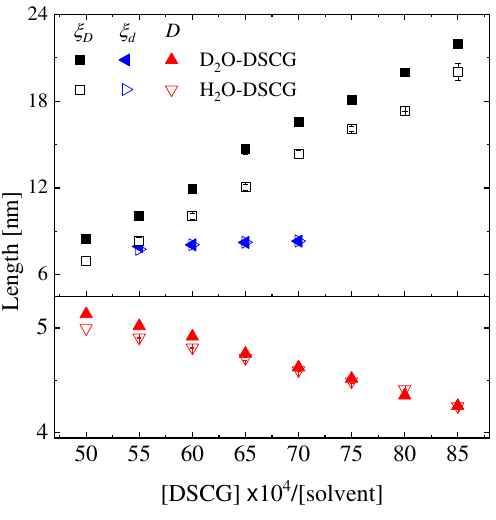}
\caption{Micro-structural information of DSCG aggregates from X-ray scattering. The inter-molecular correlation length $\xi_d$ (blue triangle), inter-aggregate correlation length $\xi_D$ (black square), and the inter-aggregate distances $D$ (red triangle) of D$_2$O-DSCG (filled symbols) and H$_2$O-DSCG (empty symbols) according to concentrations shown as molar ratio of DSCG molecules to solvent molecules. All results are obtained at $23.0 \pm 1^{\circ}$C, where both specimens are in the nematic phase.}
    \label{fig:X-ray}
\end{figure}

The wide and small angle X-ray scattering measurements~(WAXD and SAXS) show that changes in the inter-aggregate correlation length~($\xi_D$) manifest the solvent isotopic effect, \textit{i.e.}, the increase in the phase transition temperature due to D$_2$O.
As shown in Fig.~\ref{fig:X-ray}, we unveil the microstructures of DSCG aggregates, measuring the average values of inter-molecular distance~($d$), inter-molecular correlation length~($\xi_d$), inter-aggregates distance~($D$), and the inter-aggregates correlation length~($\xi_D$) using WAXD and SAXS.
We take these measurements as functions of concentrations at a fixed temperature, $23\pm1^\circ$C.
For more detailed information, refer to Fig. S1, and S2.
The $d$ remains constant at 0.34~nm, $\pi-\pi$ stacking distance, regardless of the concentrations and solvent used (not shown in Fig.~\ref{fig:X-ray}).
For both solvents, as the concentration increases, inter-molecular correlation length~$\xi_d$ slightly increases, and inter-aggregates distance~$D$ decreases. These imply that there are more aggregates that are elongated a bit.
However, the solvent change from H$_2$O to D$_2$O makes no noticeable difference in $\xi_d$ and $D$, indicating that aggregate lengths are similar.
To our interest, the inter-aggregates correlation length~$\xi_D$ of D$_2$O-DSCG is approximately 20\% longer than H$_2$O-DSCG's $\xi_D$ across all concentrations.
Note that $\xi_D$ increases considerably as the concentration and, consequently, the transition temperature increase, forming a more stable nematic phase. Namely, the longer $\xi_D$ of D$_2$O-DSCG is consistent with its higher transition temperature.

\begin{figure}[htp!]
\centering 
\includegraphics{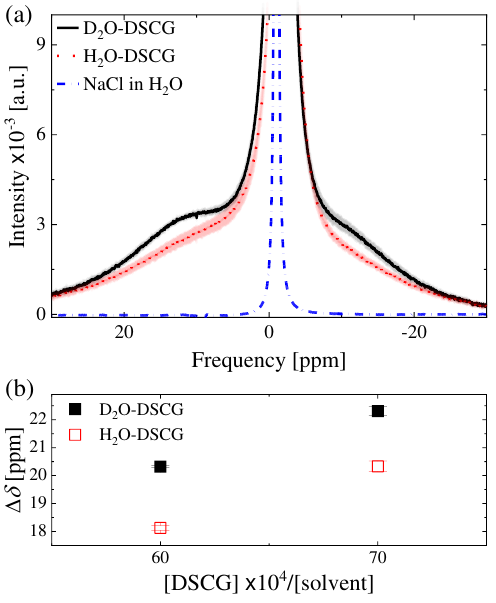}
\caption{$^{23}$Na FT-NMR intensity curve and the quadrupole splitting of nematic DSCG. (a) The $^{23}$Na FT-NMR intensity curves of D$_2$O-DSCG (black solid), H$_2$O-DSCG (red dotted), and H$_2$O-NaCl (blue dashed) solution. Each curve represents the average of multiple measurements, and the faint band around each curve corresponds to the standard deviation: three measurements for D$_2$O-DSCG, five for H$_2$O-DSCG, and one for H$_2$O-NaCl. The D$_2$O-DSCG and H$_2$O-DSCG have the same concentration, where the molar ratio [DSCG]$\times 10^4$/[solvent] = 70. For the control experiment, we use 0.5~M H$_2$O-NaCl that exhibits a single narrow peak from free Na$^{+}$ ions.
(b) Quadrupole splitting of Na$^+$ from the FT-NMR intensity curves according to the molar ratio of DSCG to solvent molecules. The error bars represent the standard deviations. See Materials and Method and Supplementary Information for curve fittings.}
    \label{fig:NMR}
\end{figure}

The X-ray scattering measurements of the microstructures disprove that replacing a hydrogen bond (H-bond) with a D-bond leads to the observed increase in the phase transition temperature of D$_2$O-DSCG. It is known that D-bonds stronger than H-bonds may affect phase behaviors of self-assembled structures, e.g., in worm-like micelles~\cite{Barron2011}. Similarly, in LCLCs, it has been reported that the increase of  H-bonds density may increase $\pi-\pi$ stacking interaction and make the aggregates stiffer and longer~\cite{Matus2017}, so we expect the stronger D-bond to affect similarly. However, we observe no difference in $\xi_d$ of D$_2$O-DSCG and H$_2$O-DSCG. Thus, we can rule out the D-bonds' effect on the phase transition temperature.

We hypothesize that solvent-induced changes in electrostatic interactions between DSCG aggregates are responsible for the change in the phase behavior. 
As shown in Fig.~\ref{fig:X-ray}, replacing H$_2$O with D$_2$O increases the inter-aggregate correlation length $\xi_D$, which reflects inter-aggregate interactions. 
We note that DSCG aggregates are negatively charged with Na$^+$ counterions, and salt-induced changes in their liquid-crystalline phase behaviors imply electrostatic interaction's roles therein~\cite{Park2008}. 
It has been reported that the lower surface charge of polyelectrolytes and resultant weaker repulsion among them stabilize their liquid crystalline phases, increasing the phase transition temperature~\cite{Stroobants1986}. 
This is because substantial charge repulsion between two rod-like aggregates favors twist deformation energetically, destabilizing the nematic phase. 
Namely, if DSCG aggregates in D$_2$O are less charged than the H$_2$O-DSCG aggregates, the less repulsion among D$_2$O-DSCG aggregates can rationalize their higher phase transition temperature in Fig.~\ref{fig:phase diagarm} and the 20\%-longer $\xi_D$ in Fig.~\ref{fig:X-ray}.

To test our hypothesis that D$_2$O-DSCG aggregates are less charged, we conduct $^{23}$Na FT-NMR spectroscopy experiments that can measure bound ion sensitively.
In Fig.~\ref{fig:NMR}(a), we compare the $^{23}$Na FT-NMR spectra of H$_2$O-NaCl, H$_2$O-DSCG, and D$_2$O-DSCG, respectively. 
The sharp peak of H$_2$O-NaCl around 0 ppm implies that Na$^+$ ions exist freely in bulk. 
In contrast, the broadening of the peaks of H$_2$O-DSCG and D$_2$O-DSCG, with satellite peaks at shoulders, originates from Na$^+$ ions bound to binding sites of the DSCG aggregates~\cite{Perahia1984, Matus2019}. 
Using a three-Lorenztian-peak model, we estimate the quadrupole splitting $\Delta\delta$, \textit{i.e.}, the separation between the satellite peak centers~\cite{Lindblom1974}, as shown Fig.~\ref{fig:NMR}(b); see Methods and Fig. S3 for details. 

The $^{23}$Na FT-NMR measurements estimate that there are indeed 10\% more bound Na$^+$ on the D$_2$O-DSCG aggregates than the H$_2$O-DSCG aggregates, implying that D$_2$O-DSCG aggregates are less charged.
The presence of anisotropic environments and the binding of Na$^+$ on the surface of aggregates results in the first-order quadrupole splitting~\cite{Lindblom1974, Goran1973}. 
The frequency difference between two split peaks, $\Delta\delta$, follows the relation 
\begin{equation}
    \Delta\delta = |S_{\mathrm{D}} \sum^n_{i=1} p_i \nu_i S_{\mathrm{M},i}|/4,
\end{equation}
where $i$ refers to different sites contributing to the quadrupole splitting. 
$S$ denotes the order parameter $\langle 3 \cos^2\theta -1 \rangle /2$ regarding an angle $\theta$, and $\langle \rangle$ means the average over the full solid angle. For $S_D$, $\theta$ is the angle between the nematic director of the DSCG and the direction of the external magnetic field for the $^{23}$Na FT-NMR measurement.
For $S_{\mathrm{M}, i}$, $\theta$ is the angle between the nematic director of the DSCG and the principal axis of the electric field gradient at the $i^{\mathrm{th}}$ site, of which the quadrupole coupling constant is $\nu_i$.
$p_i$ is the fraction of Na$^+$ ions on the $i^{\mathrm{th}}$. 
We assume that $S_{\mathrm{D}}$, $\nu_i$, and $S_{\mathrm{M}, i}$ are not affected by the solvent change from H$_{2}$O to D$_{2}$O.
This is because we conduct the $^{23}$Na FT-NMR measurements under the same conditions, and the binding sites on the aggregates remain identical.
Then, since there is one dominant binding site for Na$^+$ on the DSCG aggregates~\cite{Matus2019}, $\Delta\delta$ is proportional to the fraction of Na$^+$ bound to the site, practically reflecting the amount of the total bound Na$^+$ ion. 
To sum up, Fig.~\ref{fig:NMR}(b) approximately shows how much Na$^+$ ions are bound to the aggregates, and we conclude D$_2$O-DSCG aggregates with 10$\%$ more bound cations are less charged.

We suggest a free energy difference between H$_2$O-Na$^+$ and D$_2$O-Na$^+$ is responsible for the less charged D$_{2}$O-DSCG. 
It is reported that H$_2$O-Na$^+$ has lower free energy of solution than D$_2$O-Na$^+$ considering the librational motion of solvent molecules~\cite{Gardner1960}. Higher solubilities of Na salts in H$_2$O than in D$_2$O support this energy difference~\cite{Noonan1948, Eddy1940}.
Our experiments show that the molar solubility of sodium acetate (NaAc), the peripheral charge groups within DSCG, in H$_2$O is 4.4\% higher in H$_2$O than in D$_2$O: $\kappa_{\mathrm{H_2 O}} = 0.096 \pm 0.001$ and $\kappa_{\mathrm{D_2 O}} = 0.092 \pm 0.000$ at 295.9K.
In the same vein, the molar conductivity of H$_2$O-NaAc solution is 20\%  higher than that of D$_2$O-NaAc solution~\cite{Erickson2019}. 
These measurements semi-quantitatively support our experimental observation that H$_2$O-DSCG is 10\% more charged than D$_2$O-DSCG. Note that we exclude the possibility that the difference in the dielectric constants of H$_{2}$O and D$_{2}$O makes H$_2$O-DSCG more charged. A linear charge density is the number of charges per Bjerrum length~\cite{Park2008}, which is inversely proportional to the dielectric constant. However, H$_{2}$O and D$_{2}$O are nearly isodielectric~\cite{Wyman1938, Noonan1948}, thus the tiny difference does not explain quantitatively the 10\% charge-density difference.

\section*{Conclusion}
In summary, we investigate the solvent isotopic effect in the phase behavior of DSCG that the isotropic-to-nematic phase transition temperature of D$_2$O-DSCG is approximately 5$^\circ$C higher than the one of H$_2$O-DSCG.
Our X-ray scattering and $^{23}$Na FT-NMR measurements propose that approximately 10\% less charged D$_2$O-DSCG aggregates decrease inter-aggregate electrostatic repulsion and stabilize the nematic phase,  resulting in the increased phase transition temperature.
These findings shed new light on the solvent isotopic effects that an isotopic solvent affects counterion dissociation, the net charge of the mesogens, and the resulting lyotropic phase behavior.

\begin{acknowledgement}
The authors gratefully acknowledge the ﬁnancial support from the National Research Foundation Korea through NRF-2021R1A2C101116312. This work was also partially supported under the framework of international cooperation program managed by the National Research Foundation of Korea (NRF-2023K2A9A2A23000311)
The X-ray scattering experiments were performed at the PLS-\MakeUppercase{\romannumeral 2} 6D UNIST-PAL Beamline of Pohang Accelerator Laboratory (PAL) in the Republic of Korea (proposal number 2023-2nd-6D-A011).
UNIST Central Research Facilities (UCRF) supported FT-NMR measurements.
\end{acknowledgement}

\begin{suppinfo}

Small-angle/wide-angle X-ray scattering data and $^{23}$Na FT-NMR data.

\end{suppinfo}

\providecommand{\latin}[1]{#1}
\makeatletter
\providecommand{\doi}
  {\begingroup\let\do\@makeother\dospecials
  \catcode`\{=1 \catcode`\}=2 \doi@aux}
\providecommand{\doi@aux}[1]{\endgroup\texttt{#1}}
\makeatother
\providecommand*\mcitethebibliography{\thebibliography}
\csname @ifundefined\endcsname{endmcitethebibliography}  {\let\endmcitethebibliography\endthebibliography}{}

\end{document}


\maketitle

\renewcommand\thetable{S\arabic{table}}
\renewcommand\thefigure{S\arabic{figure}}
\renewcommand{\theequation}{S\arabic{equation}}
\renewcommand{\thepage}{S\arabic{page}}
\captionsetup{labelfont=bf}


\begin{figure}[h!]
\centering 
\includegraphics[width=1\columnwidth]{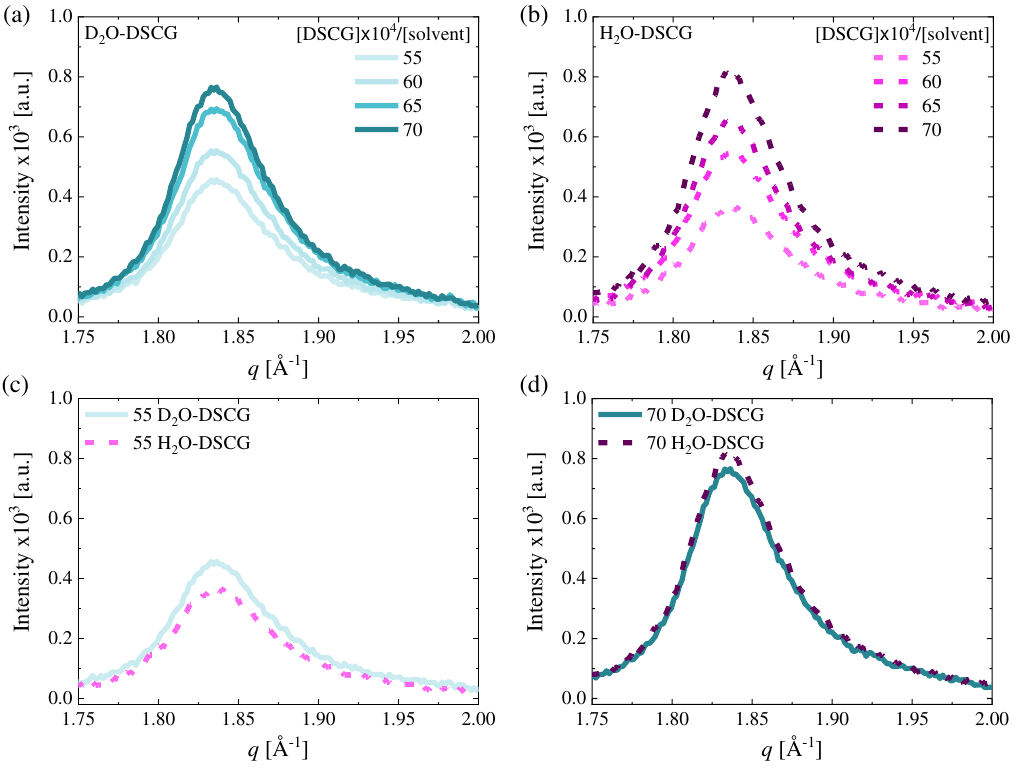}
    \caption{The background-subtracted wide-angle X-ray scattering data containing inter-molecular structural information. Scattering intensity data of (a) D$_2$O-DSCG and (b) H$_2$O-DSCG are shown according to a concentration ([DSCG]$\times 10^{4}$/[solvent] ratio) ranging from 55 to 70. The plots in (c) and (d) compare the scattering intensities of D$_2$O-DSCG and H$_2$O-DSCG at the same concentrations~([DSCG]$\times 10^{4}$/[solvent] ratio), 55 and 70, respectively.}
    \label{fig:WAXD}
\end{figure}
\par
\newpage
\begin{figure}[h!]
\centering 
\includegraphics[width=1\columnwidth]{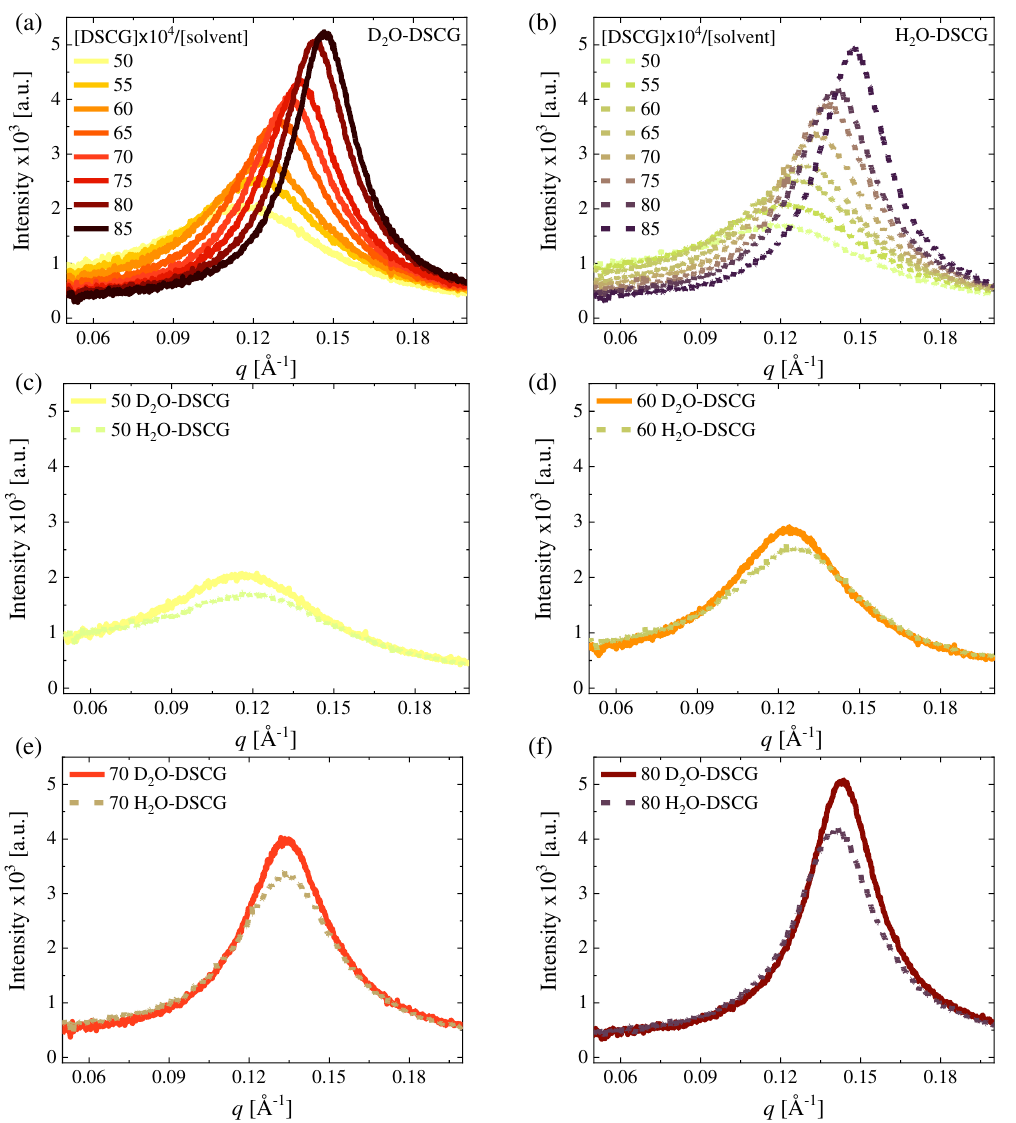}
    \caption{
    The background-subtracted small-angle X-ray scattering data containing inter-aggregate structural information. Scattering intensity data of (a) D$_2$O-DSCG and (b) H$_2$O-DSCG are shown according to a concentration ([DSCG]$\times 10^{4}$/[solvent] ratio) ranging from 50 to 85. The plots in (c - f) compare the scattering intensities of D$_2$O-DSCG and H$_2$O-DSCG at the same concentrations~([DSCG]$\times 10^{4}$/[solvent] ratio), 50, 60, 70, and 80, respectively.
    }
    \label{fig:SAXS}
\end{figure}
\newpage

\begin{figure}[h!]
\centering 
\includegraphics[width=1\columnwidth]{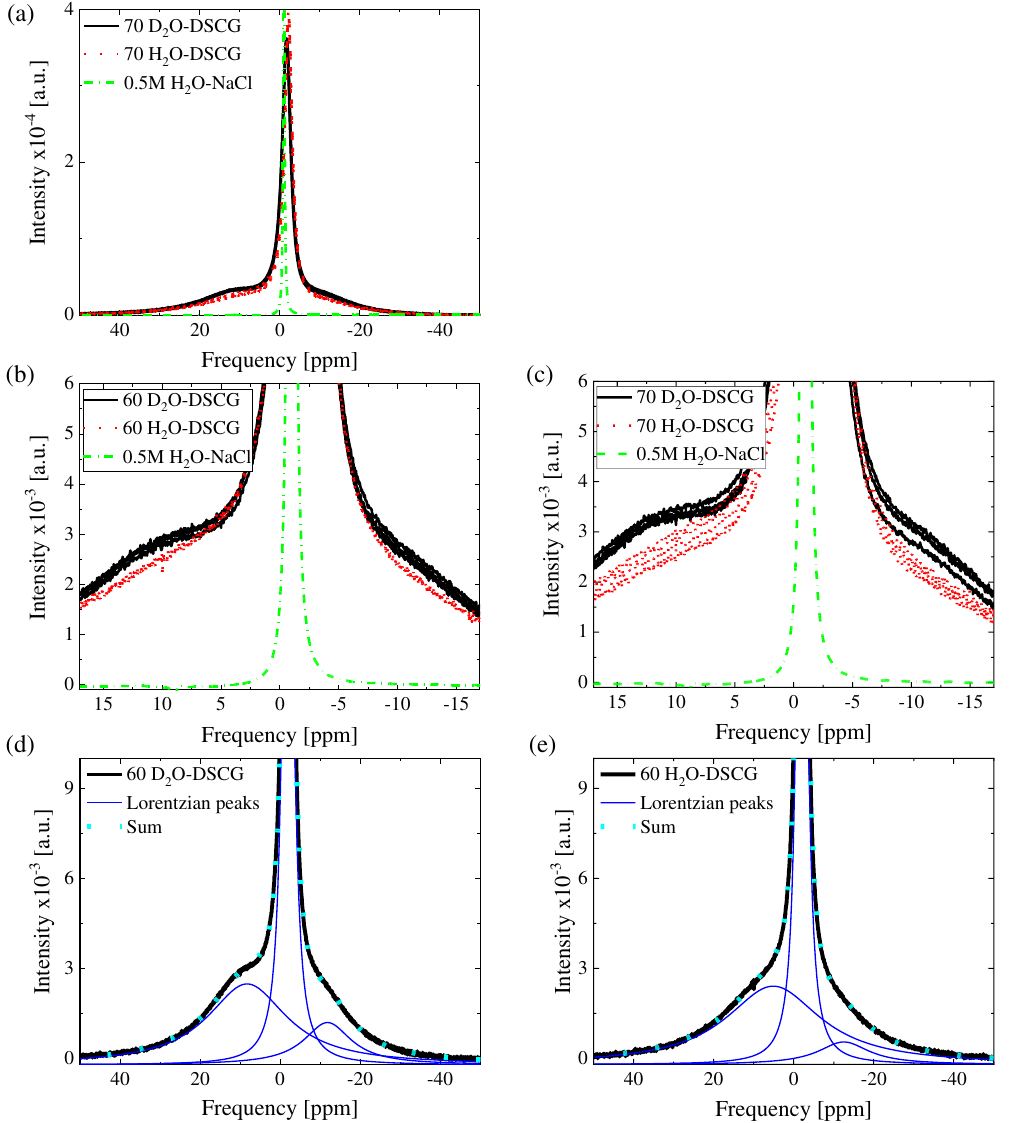}
    \caption{The $^{23}$Na FT-NMR measurements. (a) Comparison of the NMR data of D$_2$O-DSCG~(black solid) and H$_2$O-DSCG~(red dotted) at the same concentrations, [DSCG]$\times 10^4$/[solvent] = 70. The narrow peak from 0.5 M H$_2$O-NaCl is also shown. (b) and (c) Enlarged plot of the shoulder regions of the NMR data for two different concentrations: [DSCG]$\time 10^{4}$/[solvent] ratio = 60 and 70, respectively. The plots show three measurements for (b) and five for (c). (d) The three-Lorentzian peak fitting result of D$_2$O-DSCG at [DSCG]$\time 10^{4}$/[solvent] ratio = 60. The blue curves correspond to three Lorentzian curves, and the yellow dotted curve, which overlaps the NMR data, is the sum of three curves. (e) The fitting result for H$_2$O-DSCG at [DSCG]$\time 10^{4}$/[solvent] ratio = 60.}
    \label{fig:NMR}
\end{figure}

\newpage
\section*{Effects of X-ray exposure on DSCG's microstructure}
\begin{figure}[h!]
\centering 
\includegraphics{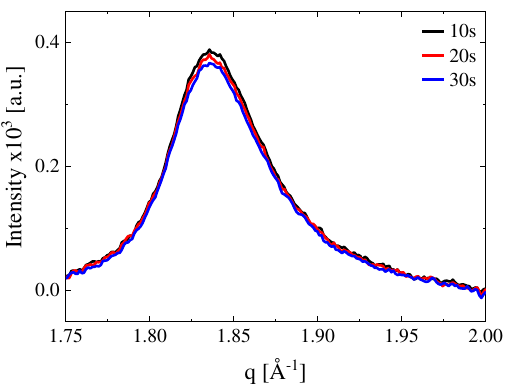}
    \caption{The background-subtracted wide-angle X-ray scattering intensities of H$_2$O-DSCG at $\mathrm{[DSCG]/[solvent]} = 55$ according to the X-ray exposure time.}
    \label{fig:x-ray damage}
\end{figure}

The background-subtracted wide-angle X-ray scattering intensities of H$_2$O-DSCG at $\mathrm{[DSCG]/[solvent]} = 55$ according to the X-ray exposure time is plotted in Fig.~\ref{fig:x-ray damage}. The decreased height of the 30-second data manifests the X-ray damage to the specimens. Nevertheless, considering the similarity between 10-s and 20-s curves, we assume that the X-ray damage bay 10-s exposure is ignorable. Note that the X-ray scattering data analyzed in Fig. 2 for the micro-structure estimation are all acquired after the exposure time of 10 seconds. 

